\documentclass[prd,a4paper]{revtex4}
\usepackage{epsfig}

\newcommand{\be}{\begin{equation}}
\newcommand{\ee}{\end{equation}}
\newcommand{\bea}{\begin{eqnarray}}
\newcommand{\eea}{\end{eqnarray}}
\newcommand{\bean}{\begin{eqnarray*}}
\newcommand{\eean}{\end{eqnarray*}}

\newcommand{\gapproxeq}{\lower
.7ex\hbox{$\;\stackrel{\textstyle >}{\sim}\;$}}
\newcommand{\lapproxeq}{\lower
.7ex\hbox{$\;\stackrel{\textstyle <}{\sim}\;$}}

\begin{document}

\bibliographystyle{unsrt}

\title{\bf Production of $f_0(1710)$, $f_0(1500)$, and $f_0(1370)$ 
in $J/\psi$ hadronic decays}

\author{Frank E. Close$^1$\footnote{e-mail: F.Close1@physics.ox.ac.uk}
and Qiang Zhao$^2$\footnote{e-mail: Qiang.Zhao@surrey.ac.uk}}

\affiliation{1) Department of Theoretical Physics,
University of Oxford, \\
Keble Rd., Oxford, OX1 3NP, United Kingdom}

\affiliation{2) Department of Physics,
University of Surrey, Guildford, GU2 7XH, United Kingdom}

\date{\today}

\begin{abstract}

A coherent study of the production of $f_0^i$ ($i=1$, 2, 3 corresponding to 
$f_0(1710)$, $f_0(1500)$, and $f_0(1370)$)  
in $J/\psi\to V f_0 \to V PP$ 
is reported based on a 
previously proposed glueball and $Q\bar{Q}$ nonet mixing scheme, 
and a factorization for the decay of $J/\psi\to V f_0^i$, where 
$V$ denotes the isoscalar vector mesons $\phi$ and $\omega$, and 
$P$ denotes pseudoscalar mesons. 
The results show that the $J/\psi$ decays are very sensitive to the 
structure of those scalar mesons, and suggest a glueball in the $1.5-1.7$ GeV region, 
in line with Lattice QCD. 
The presence of significant 
glueball mixings in the scalar wavefunctions produces 
peculiar patterns in the branching ratios for 
$J/\psi\to V f_0^i\to VPP$, which are in good agreement with 
the recently published experimental data from the BES collaboration.

\end{abstract}

\maketitle


\section{Introduction}

Although explicit evidence for a pure glueball state has never been 
confirmed in the spectroscopy of isoscalar mesons, 
experimental data are consistent with the 
scenario that a glueball should have significant mixings with quarkonia 
in the formation of some of those isoscalar mesons. This is expected to 
be especially important in the scalar mesons~\cite{scalar-meson-1}.
It is hence interesting to explore those mechanisms of 
isoscalar meson production which 
are sensitive to their glueball contents. Recently, 
Brodsky {\it et al.}~\cite{brodsky-03} proposed a possible production mechanism 
for glueball states in $e^+e^- \to \gamma^*\to J/\psi{\cal G}_J$, 
where glueball production in the same mass region as the scalar 
charmonia ($\eta_c$ and $\chi_{c0}$) would contribute to the events 
observed in $e^+e^- \to \gamma^*\to J/\psi X$ at Belle~\cite{belle}. 
In contrast to the above 
prediction for a glueball of about 3 GeV, Lattice QCD~\cite{mp,ukqcd} 
and other phenomenological studies~\cite{cfl,close-amsler,close-kirk} 
suggest the lightest glueball mass at 1.5 $\sim$ 1.7 GeV. 
Some analyses identify
the lowest $0^+$ glueball with 
the $f_0(1500)$~\cite{close-amsler,bugg,bes}, while even lower glueball masses
have been suggested in the literature~\cite{minkowski}. 
It is thus natural to concentrate on this energy region and 
explore glueball production and possible mixings 
with quarkonia in the low-lying scalars, in particular, the three 
states $f_0^i$, where $i=1$, 2 and 3 correspond to $f_0(1710)$, 
$f_0(1500)$ and $f_0(1370)$, respectively.

In Refs.~\cite{close-amsler,close-kirk}, Close {\it et al.} 
proposed a flavor mixing scheme to produce the three $f_0$ states 
among the glueball $G$, flavor nonet 
$s\bar{s}$ and $n\bar{n}$. 
It hence relates the decays  
$f_0^i\to PP$, where $P$ denotes pseudoscalar mesons 
($\pi, \ K, \ \eta, \ \eta^\prime$), 
to the mixing angles between $G$, $s\bar{s}$ and $n\bar{n}$. 
Since the mixing scheme is sensitive to the experimental 
data for $f_0^i\to PP$, 
their results not only provided a prescription for understanding the 
data from WA102~\cite{wa102},
but also highlighted the importance of the possible glueball contents 
in both $f_0(1500)$ and $f_0(1710)$, while the $f_0(1370)$ was dominated by 
the $n\bar{n}$ component. 
In this scheme, the pattern of the mass matrix will be determined 
by the relative mass positions of $G$, $s\bar{s}$ and $n\bar{n}$. 
Therefore, for the pure glueball $G$ it is crucial 
to compare with the Lattice QCD predictions. 

Independent production processes are required to examine such a
glueball-$Q\bar{Q}$ mixing mechanism and provide more decisive 
evidence for the existence of the glueball contents in some of the $f_0$ states.
At the Beijing Spectrometer (BES) 
the $f_0$ states have now been observed in both 
$J/\psi$ radiative and hadronic decays,  
and interesting signals of these scalars have been 
reported~\cite{bes-phi,bes-plb,jin}. In particular, 
the data show that these states behave contrary to expectations 
based on naive application of 
the Okubo-Zweig-Iizuka (OZI) rule~\cite{ozi} 
in $J/\psi$ decays~\cite{seiden}: 
i) the $f_0(1370)$ has been seen clearly in $J/\psi\to \phi\pi\pi$, 
but not in
$J/\psi\to \omega\pi\pi$; 
ii) there is no peak of the $f_0(1500)$ directly seen in 
$J/\psi\to \phi K\bar{K}, \ \omega K\bar{K}, \ \phi\pi\pi, \ \omega\pi\pi$, 
though its production in proton-proton scattering is quite clear~\cite{wa102};
iii) the  $f_0(1710)$ is observed clearly in both $J/\psi\to \phi K\bar{K}$ and 
$J/\psi\to \omega K\bar{K}$, but with 
$br_{J/\psi\to \omega f_0(1710)\to \omega K\bar{K}}/br_{J/\psi\to\phi f_0(1710)\to\phi K\bar{K}}\simeq 6$, 
which is against a simple $s\bar{s}$ configuration for this state. 
The  $f_0(1710)$ is not seen in $\phi \pi\pi$ and $\omega\pi\pi$, 
which is understandable due to its small branching ratio to $\pi\pi$. 
These observations hence raise questions about the configuration of the 
$f_0$ states, and also about their production mechanism in the $J/\psi$ 
hadronic decays. BES also claims observations of another scalar $f_0(1790)$, 
which is distinct from the $f_0(1710)$. We will comment 
on this state in the last Section, based on our results for the 
$f_0^{1,2,3}$.

An estimate of the low-lying glueball production rate has been made 
by extrapolating the pQCD approach of Ref.~\cite{brodsky-03} to the 
$J/\psi$ energy region~\cite{close-zhao-glueball}. 
By studying $e^+e^- \to J/\psi\to \phi G$ there, we showed that the glueball 
production rate can be approximately normalized in terms of the ideal 
$s\bar{s} (1S)$ state production. One of the results of the present 
paper will be to extract from the data the effective branching ratios 
for $J/\psi\to VG$.

In what follows, first we shall revisit the glueball-$Q\bar{Q}$ 
mixings in these three scalar $f_0^i$ mesons, 
and clarify quantitatively the sensitivity of 
the mass matrix pattern to the available experimental data 
from WA102~\cite{wa102} and BES~\cite{bes-plb,bes-phi}. 
We will then explore the implications of the glueball-$Q\bar{Q}$ 
mixings in the $f_0^i$ productions in $J/\psi\to V f_0^i$, 
and compare the results with the recent experimental data 
from BES~\cite{bes-phi,bes-plb}. 
Conclusions about the $f_0^i$ properties and their ``puzzling" 
behaviors in the $J/\psi$ decays will then be drawn.

\section{The theoretical model }

\subsection{Glueball and $Q\bar{Q}$ mixing }

References~\cite{close-amsler,close-kirk} proposed a flavour mixing scheme 
for the three scalars $f_0^{1,2,3}$ corresponding to $f_0(1710)$, 
$f_0(1500)$ and $f_0(1370)$, respectively. 
In the basis of $|G\rangle =| gg\rangle$, 
$|s\bar{s}\rangle $ and 
$|n\bar{n}\rangle \equiv |u\bar{u}+d\bar{d}\rangle/\sqrt{2}$, 
the mass matrix for the glueball-quarkonia mixing can be written as
\be
M=\left(
\begin{array}{ccc}
M_G & f & \sqrt{2} f \\
f & M_{s\bar{s}} & 0 \\
\sqrt{2} f & 0 & M_{n\bar{n}}
\end{array}
\right) \ ,
\ee
where $M_G$, $M_{s\bar{s}}$ and $M_{n\bar{n}}$ represent the masses of the pure states 
$|G\rangle$, $|{s\bar{s}}\rangle$, and $|{n\bar{n}}\rangle$, respectively.  
The parameter $f$ is the flavour independent mixing strength between 
the glueball and quarkonia, and leads to the mixture of $|G\rangle$, $|{s\bar{s}}\rangle$, 
and $|{n\bar{n}}\rangle$ to form the physical $f_0^i$ states. It can be expressed as 
$f\equiv \langle s\bar{s} | \hat{V} | G\rangle 
=\langle n\bar{n} | \hat{V} | G\rangle /\sqrt{2}$, 
where $\hat{V}$ is the potential causing such a mixing. 

Treating $f_0(1710)$, $f_0(1500)$ and $f_0(1370)$ 
as the eigenstates of the mass matrix $M$ with the eigenvalues 
of the physical resonance masses $M_1$, $M_2$ and $M_3$, respectively, 
we can then express these $f_0$ states as
\be
\left(
\begin{array}{c}
|f_0(1710)\rangle \\
|f_0(1500)\rangle \\
|f_0(1370)\rangle 
\end{array}
\right) = 
\left(
\begin{array}{ccc}
x_1 & y_1 & z_1 \\
x_2 & y_2 & z_2 \\
x_3 & y_3 & z_3 
\end{array}
\right)
\left(
\begin{array}{c}
|G\rangle \\
|{s\bar{s}}\rangle \\
|{n\bar{n}}\rangle 
\end{array}
\right) \ ,
\ee
where $x_i$, $y_i$ and $z_i$ are the mixing matrix elements determined 
by the perturbative transitions~\cite{close-amsler,close-kirk}. 

The decay of $f_0^i\to PP$ can then be factorized into contributions 
from three fundamental processes~\cite{close-kirk}:

i) the direct coupling of the quarkonia component of the three $f_0$ states 
to the final pseudoscalar meson pairs;

ii) the coupling of the glueball component of the three $f_0$ states 
to the pseudoscalar meson pairs; 

iii) the direct coupling of the glueball component to the gluonic 
component of the final state isoscalar meson pairs. 

The coupling strength of (ii) is denoted by parameter $r_2$ relative 
to the coupling of (i); analogously, parameter $r_3$ is introduced 
to account for the coupling strength of (iii) relative to (i).
In constrast to Ref.~\cite{close-kirk}, 
we make here the more natural assumption that the $G$ component
of $f_0$ states couples to the flavour singlet $Q\bar{Q}$ component 
of the $\eta$ and $\eta^\prime$. 
Taking into account the $\eta$-$\eta^\prime$ mixing, 
the factorization for $f_0^i\to PP$ is presented in Table~\ref{tab-1}.

\subsection{The production of $f_0^i$ states in $J/\psi\to V f_0^i$}

By defining the transition amplitudes via a potential $V_\phi$, 
\bea
M_{\phi G} &\equiv &\langle G | V_\phi |J/\psi\rangle \nonumber\\
M_{\phi(s\bar{s})} &\equiv &\langle s\bar{s} |V_\phi|J/\psi\rangle \nonumber\\
M_{\phi(n\bar{n})} &\equiv &\langle n\bar{n} |V_\phi|J/\psi\rangle \ ,
\eea
we can then express the transition amplitudes for $J/\psi\to \phi f_0^i$ 
in terms of the production of glueball $|G\rangle$, 
nonet $|s\bar{s}\rangle$ and $|n\bar{n}\rangle$, 
via their configuration mixings. 
Thus, the transition amplitudes can be factorized: 
\bea
M_{J/\psi\to\phi f_0^i} & = & 
\langle f_0^i|G\rangle\langle G|V_\phi|J/\psi\rangle 
+ \langle f_0^i|s\bar{s}\rangle\langle s\bar{s}|V_\phi|J/\psi\rangle 
+ \langle f_0^i|n\bar{n}\rangle\langle n\bar{n}|V_\phi|J/\psi\rangle \nonumber\\
&=& x_i M_{\phi G} + y_i M_{\phi (s\bar{s})}
+ z_i M_{\phi(n\bar{n})} \ .
\eea

\begin{figure}
\begin{center}
\epsfig{file=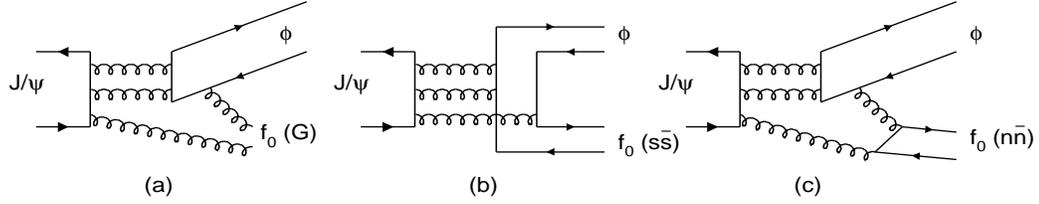, width=16cm,height=6.cm}
\caption{Production of $f_0^i$ in $J/\psi\to\phi f_0^i$. 
}
\protect\label{fig-1}
\end{center}
\end{figure}

\begin{figure}
\begin{center}
\epsfig{file=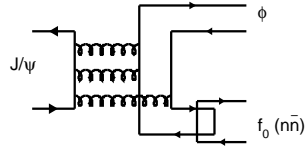, width=8cm,height=8.cm}
\caption{Process involves strong mixing of $n\bar{n}$-$s\bar{s}$ 
in the scalar production. This may occur 
at similar strength to the singly disconnected Fig.~\ref{fig-1}(b). 
}
\protect\label{fig-2}
\end{center}
\end{figure}

In Fig.~\ref{fig-1}, the leading diagrams of these transitions are illustrated. 
For $J/\psi\to \phi f_0^i$, the subprocesses (a) and (b) are of 
the same nominal order in perturbative QCD. 
Thus, we assume 
$\langle G|V_\phi |J/\psi\rangle
\simeq \langle s\bar{s}|V_\phi|J/\psi\rangle $.
The transition 
$M_{\phi(n\bar{n})}=\langle n\bar{n}|V_\phi|J/\psi\rangle$ can only occur via the doubly 
OZI disconnected process (c), which 
will be of $O(\alpha_s)$ relative to (a) and (b) 
in the perturbative regime. 
However, due to the complexity of the hadronization 
of the gluons into $Q\bar{Q}$, the role played by the doubly OZI disconnected 
diagrams is still not well-understood. 
In particular, the strong mixing of $n\bar{n}$-$s\bar{s}$ in scalar 
$0^{++}$ illustrated by the sizeable mixings (see later Section) 
shows that Fig.~\ref{fig-2} may be expected to occur at similar 
strength to the singly disconnected diagram, Fig.~\ref{fig-1}(b).
To estimate its contribution, 
we then introduce another parameter $r$, 
which describes the relative strength between the doubly OZI 
disconnected process and the singly disconnected one such that:
\be
\label{fact-amp}
\langle G|V_\phi |J/\psi\rangle
\simeq \langle s\bar{s}|V_\phi|J/\psi\rangle
\simeq \frac{1}{\sqrt{2}r}\langle n\bar{n}|V_\phi|J/\psi\rangle \ ,
\ee
where the factor $\sqrt{2}$ is from the normalization of $n\bar{n}$. 
In the above equation, we have eventually assumed that the potential 
is OZI-selecting. The left equation holds since 
$\langle G|V_\phi |J/\psi\rangle$ and 
$ \langle s\bar{s}|V_\phi|J/\psi\rangle $ both are of the same 
nominal order and both are singly OZI disconnected. 
The right equation then distinguishes the doubly OZI disconnected 
process from the singly disconnected one. 
It should be noted that contributions from 
the doubly disconnected processes with the same flavors, such as $\phi f_0(s\bar{s})$
(e.g., replace the $n\bar{n}$ in Fig.~\ref{fig-1}(c) with $s\bar{s}$), 
could also contribute in the transition 
$\langle s\bar{s}|V_\phi|J/\psi\rangle$. 
However, following the OZI-selection assumption for the 
transition potentials, such contributions can be regarded as 
being absorbed into the potential strengths. 
We will leave 
the $r$ to be determined by experimental data. Naively, $r<<1$ 
would imply that the doubly disconnected process has a perturbative 
feature, while $r \simeq 1$ could be a consistency check for the non-perturbative dominance
of $n\bar{n}$-$s\bar{s}$ mixing in the $0^{++}$ 
channel~\cite{isgur-geiger,lipkin-zou}.  

The above assumption will then allow us to express 
the partial decay width as 
\bea
\label{decay-1}
\Gamma_{J/\psi\to\phi f_0^i\to\phi PP} & = & 
\frac{|{\bf p}_{\phi i}|}{|{\bf p}_{\phi G}|}
br_{f_0^i\to PP} [  x_i + y_i  + \sqrt{2}r z_i ]^2
\Gamma_{J/\psi\to \phi G} \ ,
\eea
where ${\bf p}_{\phi i}$ and ${\bf p}_{\phi G}$ are the momenta of  
the $\phi$ meson in  $J/\psi\to \phi f_0^i$ and the virtual $J/\psi\to \phi G$, 
respectively. Their ratio gives the kinematic correction 
for the decays of $J/\psi$  
to the $\phi$ meson and states with different masses, i.e. 
$M_i\neq M_G$. 
$\Gamma_{J/\psi\to \phi G} \equiv C_0 |{\bf p}_{\phi G}| |M_{\phi G}|^2 $
is the $J/\psi$ decay width into $\phi$ and glueball $G$, 
and $|M_{\phi G}|^2$ is the invariant amplitude squared; 
$C_0$ is the phase space factor.

Similarly, the transition amplitudes for $J/\psi\to \omega f_0^i$ can be 
obtained: 
\bea
M_{J/\psi\to\omega f_0^i} & = & 
\langle f_0^i|G\rangle\langle G|V_\omega |J/\psi\rangle 
+ \langle f_0^i|s\bar{s}\rangle\langle s\bar{s}|V_\omega |J/\psi\rangle 
+ \langle f_0^i|n\bar{n}\rangle\langle n\bar{n}|V_\omega |J/\psi\rangle \nonumber\\
&=& x_i M_{\omega G} + y_i M_{\omega (s\bar{s})}
+ z_i M_{\omega (n\bar{n})} \ . 
\eea
Note that here the doubly OZI disconnected process is 
$M_{\omega (s\bar{s})}= \langle s\bar{s}|V_\omega |J/\psi\rangle $. 
Also, for the exclusive process $J/\psi\to\omega f_0^i\to \omega PP$, 
we have the partial decay width 
\bea
\label{decay-2}
\Gamma_{J/\psi\to\omega f_0^i\to\omega PP} & = & 
\frac{|{\bf p}_{\omega i}|}{|{\bf p}_{\omega G}|}
br_{f_0^i\to PP} [ x_i + r y_i + \sqrt{2} z_i ]^2
\Gamma_{J/\psi\to \omega G} \ ,
\eea
where $\Gamma_{J/\psi\to \omega G}\equiv C_0 |{\bf p}_{\omega G}| |M_{\omega G}|^2$ 
is the $J/\psi$ decay  
width into $\omega$ and pure glueball $G$; ${\bf p}_{\omega i}$ 
and ${\bf p}_{\omega G}$ represent the three momenta of the $\omega$ meson 
in $J/\psi\to\omega f_0^i$ and $J/\psi\to \omega G$, respectively. 
The flavour-blind assumption implies $|M_{\omega G}|^2=2|M_{\phi G}|^2$, 
or equivalently, $2 \Gamma_{J/\psi\to \phi G}/|{\bf p}_{\phi G}| 
=\Gamma_{J/\psi\to \omega G}/|{\bf p}_{\omega G}| $. 
For given $f_0^i$, this relation will allow us to relate the $\omega$ and $\phi$ channel 
together, and then the parameter $r$ can be determined.  


\section{Numerical results}

\subsection{Parameters and the $f_0^i$ mass matrix}

In this Section, we will quantify  these subprocesses and investigate 
the manifestation of the glueball-$Q\bar{Q}$ mixings 
in the $f_0^i$ production mechanism in the $J/\psi$ decays. 
As summarized in the Introduction, the recent data for $f_0^i$
production in $\psi$ decays exhibit apparent differences from 
the simple $Q\bar{Q}$ configurations. It is hence crucial 
for us to examine that the glueball-$Q\bar{Q}$ mixings 
obtained in the study of $f_0^i\to PP$ are consistent with the 
observations in the $J/\psi$ decays. Predictions 
for the $f_0^i$ productions in $J/\psi\to V f_0^i\to VPP$ 
will then provide a self-consistent check of this approach.

We first determine the model parameters by fitting the WA102 data~\cite{wa102}. 
Taking into account the unitarity and orthogonality constraint for the mixing matrix, 
the data can be fitted with a low reduced $\chi^2$ (see the Fit-I of 
Table~\ref{tab-2}), 
and the parameters are presented in Table~\ref{tab-3}.
Note that the $f_0^i$ decay branching ratio fractions generally 
have large errors. It shows that 
the fraction $br(f_0(1710)\to \pi\pi)/br(f_0(1710)\to K\bar{K})=0.20\pm 0.03$ 
with the relatively small errors is the major contribution to $\chi^2$. 
We also note that the constraints of unitarity and orthogonality 
will add another 9 relations in the fitting. We arbitarily 
require the error to be $3\%$. Because of this treatment, 
the absolute values of the $\chi^2$ only 
provide a reference for the fitting results.

As shown by column Fit-I in Table~\ref{tab-2}, 
the masses of the $|G\rangle$, $|s\bar{s}\rangle$, and $|n\bar{n}\rangle$ 
follow the ordering of $M_{s\bar{s}} > M_G > M_{n\bar{n}}$, 
which is the same as Ref.~\cite{close-amsler,close-kirk}.
Furthermore, in this approach we find 
that the $\eta-\eta^\prime$ mixing angle, $\phi =-18.5\pm 3.1$,
fits naturally well the expectations from linear mass relations~\cite{pdg2004}.  
It is interesting to see that the sign pattern 
of the wavefunction mixing matrix as observed in Ref.~\cite{close-kirk}
is reproduced: 
\bea
\label{mix-1}
|f_0(1710)\rangle & = & 0.39 |G\rangle + 0.91 |{s\bar{s}}\rangle + 0.13 |{n\bar{n}}\rangle \nonumber\\
|f_0(1500)\rangle & = & -0.73|G\rangle + 0.37 |{s\bar{s}}\rangle -0.57 |{n\bar{n}}\rangle \nonumber\\
|f_0(1370)\rangle & = & 0.56|G\rangle - 0.12 |{s\bar{s}}\rangle -0.82 |{n\bar{n}}\rangle  \ ,
\eea
where $|s\bar{s}\rangle$, $|G\rangle$, and $|n\bar{n}\rangle$ are dominant components
in $f_0^{1,2,3}$, respectively.

The BES experiment measured the exclusive branching ratios for 
the $f_0^i$ productions in four channels, 
$J/\psi\to \phi K\bar{K}$, $\phi\pi\pi$, $\omega K\bar{K}$ and $\omega \pi\pi$. 
In particular, according to BES~\cite{bes-plb}, 
$br(f_0(1710)\to \pi\pi)/br(f_0(1710)\to K\bar{K})$ has 
an upper limit of 0.11 with 95\% confidence level. 
This ratio, as pointed earlier, is an important contribution 
to the $\chi^2$ in the fitting of the WA102 data. 
For the $f_0(1370)$, the ratio $br(f_0(1370)\to K\bar{K})/ br(f_0(1370)\to \pi\pi)=0.08 \pm 0.08$, 
is reported, which implies a much smaller $K\bar{K}$ decay rate 
than its $\pi\pi$ decays. In comparison with the WA102 data, 
these two ratios provide additional constraints. 
For the $f_0(1500)$, we derive  
$br(f_0(1500)\to K\bar{K})/br(f_0(1500)\to \pi\pi)=(0.8\pm 0.5)/(1.7\pm 0.8)
\simeq 0.78 \pm 0.66$ by taking the bound values of the numerator 
and denominator. Although the  value overlaps with the WA102 results 
$(0.32\pm 0.07)$, the large errors will not be helpful for improving 
the $\chi^2$ in the fitting. We hence include both in the fitting 
and the parameters and results are presented 
as Fit-II in Table~\ref{tab-2} and 
Table~\ref{tab-3} to compare with Fit-I.

Again, we find the consistent $\eta-\eta^\prime$ 
mixing angle, $\phi=-22.7\pm 3.5$~\cite{pdg2004}.
The same mass ordering, $M_{s\bar{s}} > M_G > M_{n\bar{n}}$, 
and sign pattern of the mixing matrix are also preserved:
\bea
\label{mix-2}
|f_0(1710)\rangle & = & 0.36 |G\rangle + 0.93 |{s\bar{s}}\rangle + 0.09 |{n\bar{n}}\rangle \nonumber\\
|f_0(1500)\rangle & = & -0.84|G\rangle + 0.35 |{s\bar{s}}\rangle -0.41 |{n\bar{n}}\rangle \nonumber\\
|f_0(1370)\rangle & = & 0.40|G\rangle - 0.07 |{s\bar{s}}\rangle -0.91 |{n\bar{n}}\rangle  \ ,
\eea
which suggests the consistency between 
those two sets of data.

To be specific, we abstract some stable features arising from 
these two fitting results as follows:

i) Our analysis 
shows a strong preference of the $\eta$-$\eta^\prime$ mixing angle to be around 
$-23^\circ\sim -18^\circ$ degrees, which is consistent with previous 
data~\cite{pdg2004}, and should be a good reference for this factorization scheme.

ii) In the proposed mixing scheme the glueball couplings 
into isoscalar meson pairs turn out to be strong. 
This process has a strength of $r_3\sim 1$ relative to the $Q\bar{Q}$ couplings 
to the pseudoscalar meson pairs. 
This suggests the glueball-$Q\bar{Q}$ couplings in the scalar channel 
at these masses are not perturbative and cautions against a naive 
pQCD interpretation of the relative importance of Figs.~\ref{fig-1}(a), (b) 
and (c), at least for $0^{++}$ mesons.

iii) The same mass ordering, i.e. 
$M_{s\bar{s}} > M_G > M_{n\bar{n}}$, 
is found in both of these fitting results as shown 
in Table~\ref{tab-3}. This is in agreement with the results 
of Ref.~\cite{close-kirk}. The mass for the physical $f_0(1370)$ 
is treated as a free parameter, 
and we find in Fit-I that $M_3=1275\pm 34$ MeV 
and in Fit-II, $1260\pm 13$. 
The preference of low masses for the $f_0(1370)$ 
may be consistent with a recent analysis~\cite{bes-phi,bugg-comment}. 
The analyses of Ref.~\cite{bes-phi} suggest that  
the difficulty of separating the $f_2(1270)$ peak from the 
$f_0(1370)$ led to the difficulty of determining the mass pole position 
of $f_0(1370)$.
Due to the rapid rise of $f_0(1370)\to 4\pi$ phase space with mass, 
a peak occuring at about 1390 MeV in the $4\pi$ channel 
is actually farther away from the mass pole compared with 
the $\pi\pi$ channel. Refined analyses show that a pole position 
around 1315 MeV may be reasonable~\cite{bugg-comment}.

iv) In both fittings the masses of the pure $|s\bar{s}\rangle$ 
are found to be lower than the physical $f_0(1710)$, which is dominated 
by the $|s\bar{s}\rangle$ component. Furthermore, the pure 
$|n\bar{n}\rangle$ are found to be higher than the physical 
$f_0(1370)$, which is dominated by the $|n\bar{n}\rangle$. 
This suggests that the glueball mixings lead to 
a larger mass gap between $f_0(1370)$ and $f_0(1710)$ 
in comparison with the mass gap between the simple 
$|n\bar{n}\rangle$ and $|s\bar{s}\rangle$. 
This is also consistent with the observations that 
the mass gaps between the $Q\bar{Q}$ dominated axial-vectors, $f_1(1285)$ 
and $f_1(1530)$,
and between the tensors, $f_2(1270)$ and $f_2(1525)$, 
are relatively smaller than that between $f_0(1370)$ and $f_0(1710)$. 
Furthermore, the relative elevation of
the mass of the $f_0(1710)$ relative to the $f_{1,2}(\sim 1530)$ states, 
in contrast to the similar mass for the low mass states
$f_{0,1,2}(1270\sim 1370)$ suggests that the primitive glueball 
lies in the $1.5\sim 1.7$ GeV mass range (this is also in agreement with our
analysis of the scalar states, which follows).

v) The most interesting feature is the stable pattern  
of the wavefunction mixing matrix arising from these two fittings, 
which is essentially determined by the relative mass positions 
between the pure states ($|G\rangle$, $|s\bar{s}\rangle$, and $|n\bar{n}\rangle$)
and physical ones ($M_{1,2,3}$). 
Note that the mixing elements are given by 
\be
\left\{
\begin{array}{ccl}
x_i & =& (M_i-M_{s\bar{s}})(M_i-M_{n\bar{n}})C_i \\
y_i & = & (M_i-M_{n\bar{n}}) f C_i \\
z_i &= & \sqrt{2} (M_i-M_{s\bar{s}}) f C_i \ ,
\end{array}
\right.
\ee
where $C_i$ is the normalization factor of Eqs.~(\ref{mix-1}) and (\ref{mix-2}), 
and both $C_i$ and $f$ are positive. 
For the $f_0(1370)$, as an example, 
the phases can be analytically 
understood with $M_{s\bar{s}}> M_{n\bar{n}}>M_3$. 
The same analysis applies to the $f_0(1710)$, where it is essential to have 
$M_1> M_{s\bar{s}}$. For the $f_0(1500)$, 
it shows that $M_2$ can either larger or smaller than $M_G$ 
if both are located between $M_{n\bar{n}}$ and $M_{s\bar{s}}$.
This also explains the relative strengths between those components 
in the wavefunctions. In particular, 
all the components are in phase for the $f_0(1710)$; 
the $|s\bar{s}\rangle$ and $|n\bar{n}\rangle$ 
are out of phase in the $f_0(1500)$ wavefunctions (``flavor octet tendency");  
for the $f_0(1370)$, the $|n\bar{n}\rangle$ 
and $|s\bar{s}\rangle$ are in phase and hence have a ``flavor singlet tendency".
For the last, the significant $|n\bar{n}\rangle$ 
and negligible $|s\bar{s}\rangle$ 
turn out to agree with the observation that the $f_0(1370)$ 
has large branching ratios to $\pi\pi$ and $4\pi$ channels.

In brief, we find that the inclusion of the recent BES data 
does not change the phase patterns of the wavefunction mixing matrix. 
The stable phase patterns and relative strengths between the glueball and $Q\bar{Q}$ 
components can be understood as the preference of the experimental 
data to have $M_{s\bar{s}} > M_G > M_{n\bar{n}}$ with 
$1.46 < M_G < 1.52$ GeV. This turns out to be meaningful since 
such stable mixing patterns possess predictive power, which can be 
tested in independent processes.

\subsection{Constraints from the $f_0(1710)$ production }

The mixing matrices provide a starting point for considering 
the exclusive decay processes of $J/\psi\to \phi f_0^i\to \phi PP$ 
and $J/\psi\to \omega f_0^i\to \omega PP$. In particular, 
the clear signals for $f_0(1710)$ in 
$J/\psi\to \phi f_0(1710)\to \phi K\bar{K}$ and 
$J/\psi\to \omega f_0(1710)\to \omega K\bar{K}$ 
channels~\cite{bes-plb,bes-phi} will allow us 
to normalize the glueball production rates via  
Eqs.~(\ref{decay-1}) and (\ref{decay-2}) in Section IIB.
We will show in this Subsection that the 
$f_0^i$ production in $J/\psi$ decays is correlated with 
the glueball-$Q\bar{Q}$ mixings introduced in the previous Subsection, 
and a self-consistent treatment of quark-glue couplings 
is needed. We then clarify that 
the BES data for the $f_0(1710)$ will provide  not only
a crucial constraint on the model, but also a self-consistent 
examination of this approach.

From Eqs.~(\ref{decay-1}) and (\ref{decay-2}), we have 
\bea
\label{f0-1710}
\Gamma_{J/\psi\to\phi f_0^1\to\phi K\bar{K}} & = & 
\frac{|{\bf p}_{\phi 1}|}{|{\bf p}_{\phi G}|}
br_{f_0^1\to K\bar{K}} [  x_1 + y_1  + \sqrt{2}r z_1 ]^2
\Gamma_{J/\psi\to \phi G} \nonumber\\
\Gamma_{J/\psi\to\omega f_0^1\to\omega K\bar{K}} & = & 
\frac{|{\bf p}_{\omega 1}|}{|{\bf p}_{\omega G}|}
br_{f_0^1\to K\bar{K}} [ x_1 + r y_1 + \sqrt{2} z_1 ]^2
\Gamma_{J/\psi\to \omega G} \ .
\eea
The BES collaboration~\cite{bes-phi} reported 
the partial decay branching ratio 
$ \Gamma_{J/\psi\to\phi f_0^1\to\phi K\bar{K}}/\Gamma_{J/\psi}^{T}
=(2.0\pm 0.7 )\times 10^{-4}$, which is consistent with 
the estimate of the PDG~\cite{pdg2004} of $(3.6\pm 0.6)\times 10^{-4}$. 
For  
$ \Gamma_{J/\psi\to\omega f_0^1\to\omega K\bar{K}}/\Gamma_{J/\psi}^{T}$, 
BES~\cite{bes-plb} found much larger ratio $(13.2\pm 2.6) \times 10^{-4}$ than  
the PDG~\cite{pdg2004},
$(4.8\pm 1.1) \times 10^{-4}$, due to better measurement of the 
$J/\psi\to\omega K\bar{K}$ channel. 

As discussed previously, 
the coupling $r$ determines the role played by the OZI rules 
for the doubly disconnected processes. It is, in principle, 
independent of the $f_0^i$ decays, but should be determined 
by the $f_0^i$ production mechanisms in $J/\psi\to V f_0^i$. 
To see more clearly the contribution 
of the doubly OZI disconnected subprocess, we take the naive 
OZI constraint, $r\to 0$, in the perturbative limit. 
Follow the implication of the flavour-blind assumption:
$2 \Gamma_{J/\psi\to \phi G}/|{\bf p}_{\phi G}| 
=\Gamma_{J/\psi\to \omega G}/|{\bf p}_{\omega G}| $,  
and take the ratio between the $\phi$ and $\omega$ channels 
in Eq.~(\ref{f0-1710}), 
we then have
\be
R_{OZI}=\frac{\Gamma_{J/\psi\to\phi f_0^1\to\phi K\bar{K}}}
{\Gamma_{J/\psi\to\omega f_0^1\to\omega K\bar{K}}}
=\frac{| {\bf p}_{\phi 1}|}{| {\bf p}_{\omega 1} |}
\frac{[  x_1 + y_1  ]^2}{2[ x_1  + \sqrt{2} z_1 ]^2}
 \simeq 3.0 \ ,
\ee
which is contradicted by the experimental result 
$R_{exp}=0.15$ from BES~\cite{bes-plb,bes-phi} 
and $R_{exp}=0.75$ from PDG~\cite{pdg2004}. 

A solution is to 
keep the flavour-blind assumption, but 
allow $r\neq 0$, i.e. to include contributions from Fig.~\ref{fig-1}(c). 
With the experimental data for the $f_0(1710)$ 
production, 
we can determine the coupling strength $r$: 
\be
\label{ddd}
r=\frac{1}{\chi_0 y_1-\sqrt{2} z_1}
\left[(1-\chi_0)x_1 +y_1 -\sqrt{2}\chi_0 z_1\right] \ ,
\ee
where $\chi_0\equiv (2R_{exp} |{\bf p}_{\omega 1}|/|{\bf p}_{\phi 1} |)^{1/2}$.  
Note that $R_{exp}$ is determined by the experimental data; $r$ 
is not a free parameter in this calculation. 
We find that for $R_{exp}=0.15 \sim 0.75$, $r$ has a range of
$0.5\sim 2.5$, which suggests a rather strong contribution 
from the doubly OZI disconnected diagram. 
It shows that though there are significant descrepancies 
between the PDG estimates and BES measurements, 
they both favor large values for $r$ and breaking of the OZI-rule. 
In particular, this range 
is within the expectation of non-perturbative glueball-$Q\bar{Q}$ 
couplings as found in the configuration mixing scheme for 
the $f_0$ states, and as suggested by the parameters $r_2$ and $r_3$ 
in Section II. In this sense, the value $r\sim 1$
turns out to be a self-consistent result of this model.

To test the self-consistency of the calculation 
and sensitivity of other branching ratios to $r$, 
we examine the following four cases:
i) with the parameters of Fit-I, 
and $R_{exp}=0.75$ from the PDG~\cite{pdg2004}, 
we derive $r=0.5$ and the predictions are labelled as Fit-I(a); 
ii) with the parameters of Fit-I, and 
$R_{exp}=0.15$ from the BES~\cite{bes-plb,bes-phi}, 
we derive $r=2.5$ and the predictions are labelled as Fit-I(b); 
iii) with the parameters of Fit-II, 
and $R_{exp}=0.75$, we derive $r=0.6$ and the predictions are 
labelled as Fit-II(a); iv) with the parameters 
of Fit-II, and $R_{exp}=0.15$, we derive 
$r=2.2$ and the predictions are labelled as Fit-II(b). 
Recalling that Fit-II has accommodated the data from BES, 
we note in advance that Fit-II(b) is the only set where 
the new experimental data from BES have been used throughout. 
We hence would expect that predictions from Fit-II(b) are 
the self-consistent ones.

With a fixed $R_{exp}={\Gamma_{J/\psi\to\phi f_0^1\to\phi K\bar{K}}}/
{\Gamma_{J/\psi\to\omega f_0^1\to\omega K\bar{K}}}$, we are able to
normalize the glueball production widths via
Eqs.~(\ref{f0-1710}) and (\ref{ddd}). In particular, 
the spin-averaged invariant amplitudes squared are
\be
\label{glueball-norm}
|M_{\phi G}|^2=\frac 12 |M_{\omega G}|^2
=\frac{\Gamma_{J/\psi\to \phi G}}{|{\bf p}_{\phi G}|}
=\frac{\Gamma_{J/\psi\to\phi f_0^1\to\phi K\bar{K}}}
{|{\bf p}_{\phi 1}| br_{f_0^1\to K\bar{K}} 
[  x_1 + y_1  + \sqrt{2}r z_1 ]^2} \ ,
\ee
where the ratio $br_{f_0^1\to K\bar{K}}\equiv 
\Gamma(f_0(1710)\to K\bar{K})/\Gamma(f_0(1710)\to \mbox{all})$ 
will be estimated based on the present experimental information. 
So far, the experimental evaluations of $br_{f_0^1\to K\bar{K}}$ 
still have large uncertainties. Earlier analyses from Ref.~\cite{longacre} 
gave $br_{f_0^1\to K\bar{K}}=0.38$, which however may not be reliable. 
Taking into account that $K\bar{K}$ is the dominant decay mode 
for the $f_0(1710)$, and the branching ratio fractions 
of other pseudoscalar pair ($\pi\pi$, $\eta\eta $, $\eta\eta^\prime$) decays 
over $K\bar{K}$ are small~\cite{wa102,bes-phi},  
a more reliable estimate could be 
$br_{f_0^1\to K\bar{K}}\simeq 
\Gamma(f_0(1710)\to K\bar{K})/\Gamma(f_0(1710)\to 
(K\bar{K}+\pi\pi+\eta\eta +\eta\eta^\prime))\simeq 
0.60$ with the WA102 data.  
We have also assumed that the $4\pi$ decay channel is negligible~\cite{bugg-comment}.  

A large value of $br_{f_0^1\to K\bar{K}}=0.60$
will predict different glueball production rates. 
However, we would like to point out in advance that 
the magnitudes of the predicted glueball production rates
and the predicted branching ratios for the $f_0(1500)$ and $f_0(1370)$
will not change drastically with the change of 
$br_{f_0^1\to K\bar{K}}$ from 0.38 to 0.60, or even up to 0.8.
This is natural since the underlying physics in the exclusive process 
of $J/\psi\to V f_0^i$ should not be sensitive 
to the experimental uncertainties arising from the total widths of the $f_0^i$. 
We will come back to this point in the next Subsection. 
To proceed, we will then adopt $br_{f_0^1\to K\bar{K}}=0.60$  
in the calculations.

\subsection{Predictions for the glueball, $f_0(1500)$ and $f_0(1370)$ }

With  $br_{f_0^1\to K\bar{K}}=0.60$, we have access to  
the normalized invariant amplitudes squared for the glueball production
via Eq.~(\ref{glueball-norm}). The branching ratios for  
$J/\psi\to V G $ can then be derived, 
and the results are presented in Table~\ref{tab-4} 
for the four circumstances defined above.

Although there exist uncertainties due to  
dependence on the experimental data for 
$br_{f_0^1\to K\bar{K}} $, 
we find that the magnitudes are well constrained at order $10^{-4}$. 
Interestingly, the partial width of $J/\psi\to \omega G$ 
is predicted to be larger than $J/\psi\to \phi G$ by about a factor 
of two. 
A well-determined value for $R_{exp}$ will improve the 
accuracy of this prediction. Meanwhile, 
it is interesting to compare this approach with the one proposed 
in Ref.~\cite{close-zhao-glueball}. In the latter, 
the pQCD assumption provided a way to connect the electromagnetic decay 
processes with the strong decays via gluon-counting rules to the next-to-leading 
order, through which the glueball production rate can be estimated. 
However, the estimate strongly depended on the 
validity of treating all the exchanged gluons as perturbative. 
In contrast, the validity of pQCD treatment in this work can be examined 
by the factorization scheme and controlled by the experimental data.
It is worth noting, 
since the doubly OZI disconnected processes are intimately related to the 
glueball-$Q\bar{Q}$ mixings, it is not surprising that 
non-perturbative contributions are still significant in $J/\psi\to V f_0^i$.

As follows, we will examine the predictions from those four 
circumstances for the production of the $f_0(1500)$ and $f_0(1370)$, and compare 
them with the available experimental data from BES.

To calculate the partial decay widths for $J/\psi\to V f_0^i\to V PP$, 
we still need the branching ratios for $f_0^i\to PP$.  
For $f_0^i\to K\bar{K}$ and $\pi\pi$, the experimental data 
from PDG are listed in Table~\ref{tab-5}. 
Some of these values still have large uncertainties. 
However, we note that the branching ratio fractions, 
$br(J/\psi\to\phi f_0^i\to \phi PP)/br(J/\psi\to\omega f_0^i\to \omega PP)$, 
are independent of the branching ratios for $f_0^i\to PP$. 
Therefore, the theoretical predictions for the $J/\psi$ decay branching ratio 
fractions can still be compared with the experimental data. 
Again, we note that the predicted physics for the exclusive 
$J/\psi\to V f_0^i$ should not be dramatically sensitive to the 
uncertainties arising from the branching ratios for $f_0^i\to PP$.

As summarized earlier, the BES results~\cite{bes-phi} show that 
the branching ratios for $J/\psi\to \phi f_0^i\to \phi \pi\pi$ 
are sizeable for $f_0(1370)$ and $f_0(1500)$, while 
the $f_0(1710)$ is negligible. In $J/\psi\to \phi f_0^i\to \phi K\bar{K}$, 
the situation changes, and the $f_0(1710)$ has the clearest signals. 
On the other hand, signals for 
the $f_0(1370)$ and $f_0(1500)$ in $\omega \pi\pi$ and $\omega K\bar{K}$ 
are found to be small and difficult to isolate, while they are relatively 
clearer in $\phi\pi\pi$ and $\phi K\bar{K}$.

In Table~\ref{tab-6}, the branching ratios for 
the $J/\psi$ decays into $\phi K\bar{K}$, $\omega K\bar{K}$, $\phi\pi\pi$, 
and $\omega\pi\pi$ via the $f_0^i$ are presented for  Fit-I(a), Fit-I(b),
Fit-II(a), and Fit-II(b). 
It shows that Fit-I(a) and Fit-II(a) with $R_{exp}=0.75$ 
lead to sizeable branching ratios for the $f_0(1500)$ 
in $\omega\pi\pi$. Therefore, they are not favoured by the BES observations. 
This should not be surprising since BES is in favor of $R_{exp}=0.15$. 

In contrast, Fit-I(b) and Fit-II(b) 
with $R_{exp}=0.15$ 
produce consistent patterns for $f_0(1500)$ and $f_0(1370)$. 
The $f_0(1710)$ branching ratios 
in both $(\omega)\pi\pi$ and $(\phi)\pi\pi$ 
are found significantly smaller than in $(\omega)K\bar{K}$ 
and $(\phi)K\bar{K}$, respectively. 
With a large branching ratio of $(13.2\pm 2.6)\times 10^{-4}$ 
for $J/\psi\to \omega f_0(1710)\to \omega K\bar{K}$ 
as input, we obtain the corresponding ratio 
$1.45\times 10^{-4}$ for 
$J/\psi\to \omega f_0(1710)\to \omega \pi\pi$. 
This quantity seems quite big, and implies that it could be relatively 
easier to separate the $f_0(1710)$ signal in the $\omega\pi\pi$ channel 
than both $f_0(1500)$ and $f_0(1370)$. 
However, taking into account that the 
$\pi\pi/K\bar{K}$ ratio still has large uncertainties, 
e.g., BES gives an upper 
limit of 0.11 with $90\%$ CL, the branching ratio 
of $J/\psi\to \omega f_0(1710)\to \omega \pi\pi$ can be 
further suppressed with smaller $\pi\pi/K\bar{K}$ ratios.
The $f_0(1500)$ has sizeable branching ratios 
in $(\phi)\pi\pi$, while all the other channels are small, especially 
the ones in association with $\omega$ production.  
It is worth noting that an upper limit on $J/\psi\to \omega f_0(1500)$ 
is given by BES at $0.9\times 10^{-4}$~\cite{bes-f0(1500)}. Adding the branching ratios of 
$(\omega)K\bar{K}$ and $(\omega)\pi\pi$ for the $f_0(1500)$, 
the ratios are well below the limit, which is in agreement with the data.
Comparing Fit-I(b) with Fit-II(b) for the $f_0(1370)$, we find that 
the large uncertainties arising from the $K\bar{K}$ to $\pi\pi$ ratios 
do not change the major charactor of the branching ratio pattern. 
Its branching ratios in $(\phi)\pi\pi$ are significantly  
bigger than in $(\omega)\pi\pi$, which was previously regarded as 
a ``puzzle" for this state.

As manifested by Eq.~(\ref{ddd}), 
a small value for $R_{exp}$ such as 0.15, 
corresponds to $r\sim 2.2$, which implies strong non-perturbative 
contributions from the doubly disconnected processes. 
We stress again that $r$ 
is not a free parameter in this model. It is determined by the experimental 
measurement of $R_{exp}$ for the $f_0(1710)$, 
while all the other variables in Eq.~(\ref{ddd}) 
are from either kinematics or the independently-determined mixing matrix elements. 
Therefore, the ratio $R_{exp}$ also provides a good test of $r$, which 
is in a reasonable range between 0.5 to 2.2 in line with strong QCD (``non-perturbative QCD"). 

Interestingly, we also find that for much smaller $R_{exp}$, hence larger $r$,  
the production branching ratios for $f_0(1500)$ and $f_0(1370)$ 
via $\omega\pi\pi$ will increase and worsen the predictions. 
In this sense, we identify that the strong contributions from 
the doubly OZI disconnected processes are not an artificial effect. 
The preferred coupling strength is consistent with the expectation 
of a nonperturbative glueball-$Q\bar{Q}$ interaction appearing in the 
configuration mixings. 
It also suggests that an accurate measurement of $R_{exp}$ 
for the $f_0(1710)$~\cite{bes-plb,bes-phi} will be crucial for the 
study of the reaction mechanisms.

\section{Summary and perspective}

In this work, we proposed a factorization scheme 
for the study of the scalar mesons $f_0^i$ production in the 
$J/\psi$ hadronic decays into the isoscalar vector mesons $V$
($=\phi$ and $\omega$) and pseudoscalar meson pairs $PP$ 
($=\pi\pi, \ K\bar{K}, \ \eta\eta$, and $\eta\eta^\prime$), i.e.  
$J/\psi\to V f_0^i\to V PP$. 
With an improved treatment for the glueball-$Q\bar{Q}$ mixing 
in the configuration of the $f_0^i$, we found that the reaction mechanisms 
were very sensitive to the structure of the $f_0^i$. In particular, 
due to the glueball-$Q\bar{Q}$  mixing, 
the predicted branching ratios for $J/\psi\to V f_0^i\to VPP$ exhibit 
unusual patterns, which are in agreement with the experimental 
data from BES. 

The importance of glueball-$Q\bar{Q}$  mixing 
is also highlighted by the indispensible contributions from the 
doubly disconnected processes, which turn out to be nonperturbative 
and violate the OZI rule. 
Since the coupling $gg\to Q\bar{Q}$ in the doubly disconnected processes 
is essentially the same as the glueball-$Q\bar{Q}$ mixing, 
the nonperturbative feature of the doubly disconnected processes 
is self-consistent with the proposed configuration mixing scheme for these
three $f_0$ states. In this sense, our results not only provide 
an understanding of the recent ``puzzling" experimental data 
from BES~\cite{bes-phi,bes-plb,jin}, but also highlight the 
strong possibility of the existence of glueball contents 
in the $f_0(1500)$, and its sizeable interferences in $f_0(1710)$. 
Furthermore, due to the configuration mixing, 
the $|n\bar{n}\rangle$ dominant $f_0(1370)$ 
tends to have a lower mass lower than 1370 MeV, 
which also agrees with a recent more refined analysis~\cite{bes-phi,bugg-comment}. 

With Eqs.~(\ref{mix-1}) and (\ref{mix-2}), and applying 
the method of Ref.~\cite{cfl}, the  
the relative decay widths (excluding phase space) for $f_0^i\to \gamma\gamma$ are found to be
$f_0(1370):f_0(1500):f_0(1710)\sim 12:2:1$. 
The results are consistent with those of Ref.~\cite{cfl}, which should not 
be surprizing since the mixing matrices are similar to each other.

In this factorization scheme, a quantitative 
normalization of the scalar glueball production rate in the $J/\psi\to V G$ 
is also accessible. With the pure glueball mass in a range of 1.46 - 1.52 GeV, 
we obtain the branching ratios 
$br_{J/\psi\to \phi G}\simeq \frac 12 br_{J/\psi\to \omega G}\simeq (1\sim 2)\times 10^{-4}$. 
Although a direct measurement of the glueball production seems impossible, 
the success of this approach provides a key for understanding the recently 
claimed signals for another scalar $f_0(1790)$. This state is found 
distinguished from the $f_0(1710)$ based 
on two observations~\cite{bugg-comment,jin}: 
i) the $f_0(1710)$ has clear signals in $(\omega) K\bar{K}$ and $(\phi)K\bar{K}$, 
while the $f_0(1790)$ is not visible; ii) there is a definite peak 
in $\phi\pi\pi$ for the $f_0(1790)$, while the signals for $f_0(1710)$ 
is negligible. As a result, the branching ratios between $K\bar{K}$ and $\pi\pi$ 
are larger than a factor of 20. 
These distinguished features suggest that the $f_0(1790)$ may be
a radial excited state dominated by $|n\bar{n}\rangle$ which appears not to be strongly mixed
with the ground state.

As pointed out in Ref.~\cite{close-ichep04}, the decays
$J/\psi\to \gamma f_0\to \gamma\gamma V$, where $V=\omega$, $\phi$, and 
$\rho^0$, can provide further independent information about the $f_0$ states. 
Using the discussions of Ref.~\cite{cfl}, we find the widths for 
$J/\psi\to\gamma f_0^i$ satisfy 
$f_0(1710)>f_0(1500)>f_0(1370)$, which is also consistent with the BES 
data.  
Further improved
statistics at BES for these $f_0$ states in both hadronic and radiative 
decays should be able to quantitize 
the model predictions, and provide us with more information 
about their structures.

\section*{Achowledgement}

This work is supported,
in part, by grants from
the U.K. Engineering and Physical
Sciences Research Council Advanced Fellowship (Grant No. GR/S99433/01),
and
the Particle Physics and
Astronomy Research Council, and the
EU-TMR program ``Eurodice'', HPRN-CT-2002-00311.
The authors are in debt to D.V. Bugg 
for many useful comments on an early version 
of this manuscript. Useful discussions with B.S. Zou  and S. Jin, 
and comments from C.P. Shen and C.Z. Yuan,  
are also acknowledged.


\begin{table}[ht]
\begin{tabular}{lcl}
\hline
$\gamma^2(f_0^i\to \eta\eta)$ 
& & $[\alpha^2 z_i +\sqrt{2} \beta^2 y_i 
+r_2 x_i + (\sqrt{2} \alpha -\beta)^2 r_3 x_i]^2$ 
\\[1ex]
$\gamma^2(f_0^i\to \eta\eta^\prime)$ 
& & $2  [\alpha\beta (z_i-\sqrt{2}y_i) 
+ (\sqrt{2} \alpha -\beta)(\sqrt{2}\beta+\alpha) r_3 x_i ]^2$ \\[1ex]
$\gamma^2(f_0^i\to \pi\pi)$ 
& & $3 [z_i +r_2 x_i]^2$
\\[1ex]
$\gamma^2(f_0^i\to K\bar{K})$
& & $4[\frac 12 (z_i +\sqrt{2} y_i) +r_2 x_i]^2$
\\[1ex]
\hline
\end{tabular}
\caption{The theoretical reduced partial widths. Parameters 
$\alpha$ and $\beta$ are defined as 
$\alpha\equiv (\cos\phi-\sqrt{2}\sin\phi)/\sqrt{3}$ and 
$\beta\equiv (\sin\phi+\sqrt{2}\cos\phi)/\sqrt{3}$, where 
$\phi$ is the flavour octet-singlet mixing angle between 
$\eta$ and $\eta^\prime$. Factor $\gamma^2$ denotes the 
invariant decay couplings for $f_0^i\to PP$. We adopt the same 
form factors as Ref.~\cite{close-amsler,close-kirk} in the calculations.}
\label{tab-1}
\end{table}

\begin{table}[ht]
\begin{tabular}{l|cc|cc||c|cc}
\hline
B.R. fractions    & & Data from WA102 &  Fit-I &  & Data of WA102 plus BES & Fit-II & \\[1ex]
\hline
$ \ \ \ \ D $
&&  ${\bf 2.17 \pm 0.9 }$ &{ 2.34 }& &    &   &\\
$ \ \ \ \ 1/D$ && & && ${\bf 0.08 \pm 0.08 }  $ &  { 0.07 } &\\[1ex]
$\frac{f_0(1370)\to \eta\eta}{f_0(1370)\to K\bar{K}}$
&& $0.35\pm 0.21$ & 0.62 & & $0.35\pm 0.21$ & 0.17 & \\[1ex]
$\frac{f_0(1500)\to \pi\pi }{f_0(1500)\to \eta\eta}$
&& $5.5\pm 0.84 $ & 4.59 & & $5.5\pm 0.84 $ & 2.92 & \\[1ex]
$\frac{f_0(1500)\to K\bar{K}}{f_0(1500)\to \pi\pi}$
&& ${\bf 0.32 \pm 0.07 }$ & {  0.41  }&& $ 0.32 \pm 0.07 $ & {  0.42} & \\
&& & && ${\bf 0.78 \pm 0.66 }$ & & \\[1ex]
$\frac{f_0(1500)\to \eta\eta^\prime}{f_0(1500)\to \eta\eta}$
&& $ 0.52 \pm 0.16 $ & 0.68 & & $ 0.52 \pm 0.16 $ & 0.73 & \\[1ex]
$\frac{f_0(1710)\to \pi\pi}{f_0(1710)\to K\bar{K}}$
&& ${\bf 0.20 \pm 0.03 }$ & {  0.19 }& & $<{\bf  0.11 }$  & {  0.13 } & \\[1ex]
$\frac{f_0(1710)\to \eta\eta}{f_0(1710)\to K\bar{K}}$
&& $0.48\pm 0.14$ & 0.22 & & $0.48\pm 0.14$ & 0.20 & \\[1ex]
$\frac{f_0(1710)\to \eta\eta^\prime}{f_0(1710)\to \eta\eta}$
&& $<0.05 \ (90\% cl)$ & 0.03 & & $<0.05 \ (90\% cl)$ & 0.03  & \\[1ex]
\hline
 \ \ \ $\chi^2$ && & 4.6 & & & 3.6 & \\[1ex]
\hline
\end{tabular}
\caption{Fitting results in comparison with the data from WA102 and BES. 
Quantity $D$ denotes $ \frac{f_0(1370)\to \pi\pi}{f_0(1370)\to K\bar{K}}$. 
The differences between these two experimental sets applied for the
fittings are highlighted by the bold font. }
\label{tab-2}
\end{table}

\begin{table}[ht]
\begin{tabular}{l|cc||cc}
\hline
Parameters && Fit-I && Fit-II \\[1ex]
\hline
$\phi$ (degree) && $-18.5 \pm 3.1$ && $-22.7 \pm 3.5$ \\[1ex]
$M_G$ (MeV) &&  $1464\pm 47$ && $1519\pm 41$ \\[1ex]
$M_{s\bar{s}}$ (MeV) &&  $1674\pm 10$ && $1682\pm 4$ \\[1ex]
$M_{n\bar{n}}$ (MeV) &&  $1357\pm 23$ && $1304\pm 7$ \\[1ex]
$f$ (MeV) &&  $84\pm 14$ && $71\pm 5$  \\[1ex]
$r_2$  && $1.10\pm 0.02$ && $0.88\pm 0.12$  \\[1ex]
$r_3$  && $1.12\pm 0.27$ && $1.08\pm 0.20$ \\[1ex]
$M_3$ (MeV) && $1275\pm 34$ && $1260\pm 13$ \\[1ex]
\hline
\end{tabular}
\caption{Parameters for the two sets of fittings. }
\label{tab-3}
\end{table}

\begin{table}[ht]
\begin{tabular}{l||cc|cc|cc|cc}
\hline
Br. for $G$ ($\times 10^{-4}$) && Fit-I(a) && Fit-I(b) && Fit-II(a) && Fit-II(b) \\[1ex]
\hline
$br_{J/\psi\to\phi G}$ && 3.97 && 1.36 && 3.96 && 1.63 \\[1ex]
$br_{J/\psi\to\omega G}$ && 9.02 && 3.09 && 9.11 && 3.75 \\[1ex]
\hline
\end{tabular}
\caption{Normalized branching ratios for the glueball production in $J/\psi\to \phi G$ and 
$J/\psi\to \omega G$. The four sets of calculations correspond to 
Fit-I(a): with the parameters fitted by Fit-I, 
and $R_{exp}=0.75$ from the PDG~\cite{pdg2004};
Fit-I(b): with the parameters fitted by Fit-I, and 
$R_{exp}=0.15$ from the BES~\cite{bes-plb,bes-phi};
Fit-II(a): with the parameters fitted by Fit-II, 
and $R_{exp}=0.75$;
and Fit-II(b): with the parameters 
fitted by Fit-II, and $R_{exp}=0.15$.
We note that Fit-II(b) is the one 
treating the experimental data self-consistently. }
\label{tab-4}
\end{table}

\begin{table}[ht]
\begin{tabular}{l|ccc}
\hline
$br_{f_0^i\to PP}$ && $K\bar{K}$ & $\pi\pi$ \\[1ex]
\hline
$f_0(1710)$ && 0.60 & $0.60\times 11\%$ \\[1ex]
$f_0(1500)$ && 0.086 & 0.349  \\[1ex]
$f_0(1370)$ && $0.20\times 10\% $ & 0.20 \\[1ex]
\hline
\end{tabular}
\caption{Experimental values for the branching ratios of $f_0^i\to PP$. 
For $ br_{f_0(1710)\to \pi\pi}$ and $br_{f_0(1370)\to K\bar{K}}$, 
we apply the results from the BES Collaboration~\protect\cite{bes-plb,bes-phi}. }
\label{tab-5}
\end{table}

\begin{table}[ht]
\begin{tabular}{cc|cc|cc|cc|cc}
\hline
Fit-I(a) && $K\bar{K} \ (\phi)$  & Data
& $K\bar{K} \ (\omega)$  &  Data & $\pi\pi \ (\phi)$ & Data & $\pi\pi \ (\omega)$ & \\[1ex]
\hline
$f_0(1710) \ (\times 10^{-4})$ && {\bf 3.60 } & [3.6$\pm$ 0.6] & {\bf 4.80 } & [4.8$\pm$ 1.1] & 0.40 & - & 0.53 & 
- \\[1ex]
$f_0(1500) \ (\times 10^{-4})$ && 0.19 & (0.8$\pm$ 0.5)& 1.38 & - & 0.77 &  (1.7$\pm$ 0.8) & 5.60 & -\\[1ex]
$f_0(1370) \ (\times 10^{-4})$ && 0.00 & (0.3$\pm$ 0.3) & 0.08 & -  & 0.01 & (4.3$\pm$ 1.1) & 0.83 & -\\[1ex]
\hline
Fit-I(b) && $K\bar{K} \ (\phi)$  & Data
& $K\bar{K} \ (\omega)$  &  Data & $\pi\pi \ (\phi)$ & Data & $\pi\pi \ (\omega)$ & \\[1ex]
\hline
$f_0(1710) \ (\times 10^{-4})$ && {\bf 2.00 } & (2.0$\pm$ 0.7) & {\bf 13.20 } & (13.2$\pm$ 2.6) & 0.22 & - & 1.45 
& - \\[1ex]
$f_0(1500) \ (\times 10^{-4})$ && 0.66 & (0.8$\pm$ 0.5)& 0.09 & - & 2.67 &  (1.7$\pm$ 0.8) & 0.35 & -\\[1ex]
$f_0(1370) \ (\times 10^{-4})$ && 0.19 & (0.3$\pm$ 0.3) & 0.05 & -  & 1.93 & (4.3$\pm$ 1.1) & 0.54 & -\\[1ex]
\hline
\hline
Fit-II(a) && $K\bar{K} \ (\phi)$  & Data
& $K\bar{K} \ (\omega)$  &  Data & $\pi\pi \ (\phi)$ & Data & $\pi\pi \ (\omega)$ & \\[1ex]
\hline
$f_0(1710) \ (\times 10^{-4})$ && {\bf 3.60 } & [3.6$\pm$ 0.6] & {\bf 4.80 } & [4.8$\pm$ 1.1] & 0.40 & - & 0.53 & 
- \\[1ex]
$f_0(1500) \ (\times 10^{-4})$ && 0.23 & (0.8$\pm$ 0.5)& 1.20 & - & 0.92 &  (1.7$\pm$ 0.8) & 4.85 & -\\[1ex]
$f_0(1370) \ (\times 10^{-4})$ && 0.01 & (0.3$\pm$ 0.3) & 0.18 & -  & 0.14 & (4.3$\pm$ 1.1) & 1.81 & -\\[1ex]
\hline
Fit-II(b) && $K\bar{K} \ (\phi)$  & Data
& $K\bar{K} \ (\omega)$  &  Data & $\pi\pi \ (\phi)$ & Data & $\pi\pi \ (\omega)$ & \\[1ex]
\hline
$f_0(1710) \ (\times 10^{-4})$ && {\bf 2.00 } & (2.0$\pm$ 0.7) & {\bf 13.20 } & (13.2$\pm$ 2.6) & 0.22 & - & 1.45 
& - \\[1ex]
$f_0(1500) \ (\times 10^{-4})$ && 0.46 & (0.8$\pm$ 0.5)& 0.13 & - & 1.89 &  (1.7$\pm$ 0.8) & 0.52 & -\\[1ex]
$f_0(1370) \ (\times 10^{-4})$ && 0.26 & (0.3$\pm$ 0.3) & 0.09 & -  & 2.63 & (4.3$\pm$ 1.1) & 0.94 & -\\[1ex]
\hline
\end{tabular}
\caption{Predictions for $br_{J/\psi\to V f_0^i\to  V PP}$ in comparison 
with the available experimental data. The BES results~\cite{bes-plb,bes-phi} are quoted 
in the round brackets, while the PDG results~\cite{pdg2004} are
quoted in the square brackets. The symbol `-' means 
that signals of the corresponding states have not been observed in BES experiment.
See the caption of Table~\ref{tab-4} for definition of these four sets. }
\label{tab-6}
\end{table}


\end{document}